\documentclass[aps,pra,preprint,tightenlines,showpacs,groupedaddress]{revtex4}
\usepackage{graphicx}
\usepackage{amsmath}
\usepackage{longtable}
\usepackage{afterpage}
\usepackage{rotating}

\begin{document}

\title{Blackbody radiation shift of B$^+$ clock transition}
\author{Yongjun Cheng$^{1,2}$ and J. Mitroy$^{1}$ } 
\affiliation {$^1$School of Engineering, Charles Darwin University, Darwin NT 0909, Australia} 
\affiliation {$^2$Center for Theoretical Atomic and Molecular Physics, The Academy of Fundamental and Interdisciplinary Science, Harbin Institute of Technology, Harbin 150080, People$'$s Republic of China}

\date{\today}

\begin{abstract}
A calculation of the blackbody radiation shift of the B$^+$ clock 
transition is performed. The polarizabilities of the B$^+$ 
$2s^2$ $^1$S$^e$, $2s2p$ $^1$P$^o$, and $2s2p$ $^3$P$^o$ states 
are computed using the configuration interaction method with an 
underlying semi-empirical core potential. The recommended dipole 
polarizabilities are 9.64(3) $a_0^3$, 7.78(3) $a_0^3$ and 16.55(5) 
$a_0^3$ respectively. The derived frequency shift for the $2s^2$ 
$^1$S$^e$ $\to$ $2s2p$ $^3$P$^o_0$ transition at 300 K is 0.0160(5) Hz. 
The dipole polarizabilities agree with an earlier relativistic 
calculation (Safronova {\em et al.} Phys.~Rev.~Lett. {\bf 107} 143006 (2011))
to better than 0.2$\%$. Quadrupole and octupole polarizabilities 
and non-adiabatic multipole polarizabilities are also reported.    

\end{abstract}


\maketitle

\section{Introduction} 

Recent advances in laser control of atoms and ions have lead
to major improvements in the precision of optical frequency 
standards \cite{gill05a,margolis09a}. These improvements are 
expected to result in a new definition of the second \cite{gill11a}.  
Indeed, an optical clock using the Al$^+$ ion using quantum logic 
technology has been developed with a fractional frequency uncertainty 
of $8.6 \times 10^{-18}$ \cite{chou10a}. This uncertainty is 
equivalent to a drift of 1 second in $3.7 \times 10^9$ years.  

The ultra-high precision achieved by these optical frequency 
standards means they are sensitive to very small environmental 
influences. One of these influences is the blackbody 
radiation (BBR) emitted by the apparatus in which the 
atomic or ionic clock is enclosed. This BBR radiation, 
by means of the AC Stark effect changes the energies of the two 
states of the clock transition, and this can alter the 
frequency of the atomic clock \cite{margolis03a,mitroy10a}. 

This BBR-shift is in principle one of the largest sources of systematic 
error in these clocks \cite{mitroy10a,margolis03a,palchikov03a,
porsev06b,zelevinsky07a,itano07a,rosenband06a}. The BBR-shift 
(in Hz) can be written 
\begin{equation}
\Delta \nu_{\rm BBR} = 6.579684 \times 10^{15} \left(
\Delta E_{\rm upper} - \Delta E_{\rm lower} \right),
\label{BBR3}
\end{equation}
where the electric dipole (E1) induced BBR energy shift of an 
atomic state can be approximately calculated as \cite{porsev06a}
\begin{equation}
\Delta E \approx -\frac{2}{15} (\alpha \pi )^3 \alpha_1(0) T^4  \ . 
\label{BBR1}
\end{equation}
The dipole polarizability of the relevant quantum state is $\alpha_1$  
and $T$ is the temperature. Knowledge of the dipole polarizabilities 
permits a temperature dependent BBR correction to be made to the clock.  
The uncertainty in the E1 BBR shift can be written 
\begin{equation}
\delta (\Delta \nu_{\rm BBR}) = \Delta \nu_{\rm BBR} \left( \frac{\delta (\Delta \alpha_1)} {\Delta \alpha_1}
+ \frac{4 \delta T} {T} \right).
\end{equation}

Calculations of the B$^+$ ($2s^2$ $^1$S$^e_0$ - $2s2p$ $^3$P$^{o}_0$) clock 
transition have previously been made \cite{safronova11c} using a relativistic 
configuration interaction (CI) calculation to account for valence correlations 
while an all-order many-body perturbation theory approach is used to account 
for core and core-valence correlations. The paper reported the dipole 
polarizabilities and demonstrated that the clock transition for this ion 
had a relatively small 300 K BBR shift of 0.0159 Hz. The present manuscript 
confirms this result and extends the dataset for B$^+$ to encompass higher 
order polarizabilities. Calculations are performed using the CI method 
with a semi-empirical core-polarization potential to encompass core-valence 
correlations.  

\section{Methodology} 

The CI calculations used to generate the physical and $L^2$ pseudo
states were similar in style to those used previously to determine
the dispersion parameters and polarizabilities of a number of two
electron systems \cite{mitroy03f,mitroy04b,mitroy08k,mitroy09b}.
The Hamiltonian for the two active electrons is written
\begin{eqnarray}
H  &=&  \sum_{i=1}^2 \left(  -\frac {1}{2} \nabla^2_i
 + V_{\rm dir}({\bf r}_i) + V_{\rm exc}({\bf r}_i) +  V_{\rm p1}({\bf r}_i) \right) \nonumber \\
 &+&  V_{\rm p2}({\bf r}_1,{\bf r}_2) + \frac{1}{r_{12}} \ .
\end{eqnarray}
The direct, $V_{\rm dir}$, and exchange, $V_{\rm exc}$, interactions of
the valence electrons with the Hartree-Fock (HF) core were calculated 
exactly. The $1s^2$ core wave function was taken from a HF calculation 
of the B$^{2+}$ ground state using a Slater type orbital (STO) basis. 
The $\ell$-dependent polarization potential, $V_{\rm p1}$, was 
semi-empirical in nature with the functional form
\begin{equation}
  V_{\rm p1}({\bf r})  =  -\sum_{\ell m} \frac{\alpha_{\rm core} g_{\ell}^2(r)}{2 r^4}
           |\ell m \rangle \langle \ell m| .
\label{polar1}
\end{equation}

The coefficient, $\alpha_{\rm core}$, is the static dipole polarizability 
of the core and $g_{\ell}^2(r) = 1-\exp \bigl($-$r^6/\rho_{\ell}^6 \bigr)$
is a cutoff function designed to make the polarization potential finite
at the origin. The cutoff parameters, $\rho_{\ell}$, were tuned to
reproduce the binding energies of the B$^{2+}$ $ns$ ground state and
the $np$, $nd$ and $nf$ excited states. The core polarizability
was chosen to be $\alpha_{\rm core} = 0.019644$ $a_0^3$ \cite{jiang11a, bhatia97a}.
The cutoff parameters for $\ell = 0 \to 3$ were 0.6835, 0.6899, 0.8874 
and 2.945 $a_0$ respectively.

To get more accurate energy levels and polarizabilities, it is essential 
to include a two body polarization term, $V_{p2}$, in the Hamiltonian.  
The polarization of the core by one electron is influenced by the 
presence of the second valence electron. Omission of the two-body 
term would typically result in a $2s^2$ state that would be too tightly 
bound. A discussion of the importance of the two body polarization potential 
can be found in \cite{norcross76a}. The two body polarization potential is
adopted in the present calculation with the form 
\begin{equation}
V_{\rm p2}({\bf r}_i,{\bf r}_j) = -\frac{\alpha_d} {r_i^3 r_j^3}
({\bf r}_i\cdot{\bf r}_j)g_{\rm p2}(r_i)g_{\rm p2}(r_j)\ ,
\label{polar2}
\end{equation}
where $g_{p2}$ has the same functional form as $g_{\ell}(r)$.
The cutoff parameter for $g_{\rm p2}(r)$ was chosen as 0.6867 $a_0$, 
the average of $\rho_0$ and $\rho_1$ (the $\rho_2$ and $\rho_3$ 
cutoff parameters are influenced by finite nuclear mass effects, 
and thus they were not used in determining the cutoff parameter 
for $V_{\rm p2}$). Use of 0.6867 $a_0$ for the two-body cutoff 
parameter resulted in energies that were close to the experimental 
binding energies for most of the lowest lying states of B$^+$.  
Some small adjustments to the $\rho_{\ell}$, described later, were 
made later to further improve agreement with the experimental B$^+$ 
spectrum. The approach to solve the Schrodinger equation is 
termed as configuration interaction plus core polarization (CICP).   

There were a total of 163 valence orbitals with a maximum orbital
angular momentum of $\ell = 5$. The radial dependence of the orbitals 
were described by a mixture of STOs and Laguerre type orbitals
(LTOs) \cite{mitroy03f}. The number of active orbitals for 
$\ell = 0 \to 5$ were 32, 32, 30, 25, 25, and 19 respectively.
Some $\ell = 0$ valence orbitals were generated from the STOs 
used for the core. All the other orbitals were written as LTOs 
due to their superior linear dependence properties when compared 
with STO basis sets. The use of the large orbital basis resulted 
in wave functions and energies for the low-lying states that 
were close to convergence.  

The length of the CI expansions for the different states of B$^+$ 
ranged from 2000-5000. Some small changes were made to the $\rho_\ell$ 
values that were originally tuned to the B$^{2+}$ spectrum to improve 
the agreement of the B$^{+}$ energies with experiment. The oscillator 
strengths were computed with operators that included polarization 
corrections \cite{hameed72a,mitroy93a,mitroy03f}. The cutoff parameter 
in the polarization correction to dipole operator was 0.6867 $a_0$.

\section{Results and Discussion}

\subsection{Energy levels} 

The energy levels of the present calculations are given in Table 
\ref{energy} and compared with experiment. The biggest discrepancy
for the B$^{2+}$ ion was 10$^{-4}$ a.u.. The cut-off parameters
of the the polarization potential were tuned to reproduce the 
experimental binding energies of the lowest states of each symmetry.

Small adjustments to the cut-off parameters were made for the
calculations of the B$^+$ states. For example, the value of $\rho_0$
was reset to 0.7064 $a_0$ for the calculation of the states of
the $^1$S$^e$ symmetry. The value of $\rho_0$ was fixed by requiring
that the theoretical and experimental energies for the $2s^2$ state
be the same. Other fine tunings of the cut-off parameters were
made for all symmetries. The biggest discrepancy between theoretical
and experimental energies occurs for the $^1$S$^e$ symmetry and
is only $2 \times 10^{-4}$ a.u..

\begin{table}[t]
\caption[]{ \label{energy}
Theoretical and experimental energy levels (in Hartree) for some of 
the low-lying states of the B$^+$ and B$^{2+}$ ions. The energies 
are given relative to the energy of the B$^{3+}$ core. The experimental 
energies for the multiplet states are averages with the usual 
$(2J+1)$ weighting factors. The experimental data were taken from 
the National Institute of Standards and Technology \cite{nistasd500}. }  

\begin{ruledtabular}
\begin{tabular}{lcc}
State & Present & Experiment     \\ 
\hline
 \multicolumn{3}{c}{B$^{2+}$}      \\
$2s$ $^2$S$^e$ &  $-$1.393924   & $-$1.393924 \\
$2p$ $^2$P$^o$ &  $-$1.173483   & $-$1.173483 \\
$3s$ $^2$S$^e$ &  $-$0.572792   & $-$0.572863  \\
$3p$ $^2$P$^o$ &  $-$0.514642   & $-$0.514743   \\
$3d$ $^2$D$^e$ &  $-$0.500561   & $-$0.500561   \\
$4s$ $^2$S$^e$ &  $-$0.310856   & $-$0.310891  \\
$4p$ $^2$P$^o$ &  $-$0.287444   & $-$0.287498   \\
$4d$ $^2$D$^e$ &  $-$0.281527   & $-$0.281529  \\
$4f$ $^2$F$^o$ &  $-$0.281269   & $-$0.281269  \\
\multicolumn{3}{c}{B$^{+}$}   \\ 
$2s^2$ $^1$S$^e$ & $-$2.318347  & $-$2.318347     \\
$2s2p$ $^3$P$^o$ & $-$2.148168  & $-$2.148168    \\
$2s2p$ $^1$P$^o$ & $-$1.983927  & $-$1.983927    \\
$2p^2$ $^3$P$^e$ & $-$1.867605  & $-$1.867605    \\
$2p^2$ $^1$D$^e$ & $-$1.851947  & $-$1.851947    \\
$2p^2$ $^1$S$^e$ & $-$1.736606  & $-$1.736679    \\
$2s3s$ $^3$S$^e$ & $-$1.727053  & $-$1.727053    \\
$2s3s$ $^1$S$^e$ & $-$1.691092  & $-$1.691293    \\
$2s3p$ $^3$P$^o$ & $-$1.662237  & $-$1.662269     \\
$2s3p$ $^1$P$^o$ & $-$1.661828  & $-$1.661765     \\
$2s3d$ $^3$D$^e$ & $-$1.631961  & $-$1.631961     \\
$2s3d$ $^1$D$^e$ & $-$1.613484  & $-$1.613545     \\
\end{tabular}
\end{ruledtabular}
\end{table}

The agreement between the theoretical and experimental energy
levels is sufficiently close to discount the possibility that 
energy level considerations might make a significant contribution
to the uncertainty in the radial matrix elements.

\subsection{Oscillator strengths of low-lying transitions}

\begingroup
\squeezetable 
\begin{table*}[tbh]
\caption[]{ \label{ostrength}
Absorption oscillator strengths for various dipole transition lines 
of the B$^{+}$ and B$^{2+}$ ions. The experimental energy differences were 
used in the calculation of the CICP oscillator strengths.}
\begin{ruledtabular}
\begin{tabular}{lccccccc}
Transition  & CICP &BCICP & MCHF-BP & MCHF & CI  & Other Theory & Experiment \\ 
\hline  
          \multicolumn{8}{c}{B$^{2+}$}   \\
 $2s$$\to$$2p$ & 0.36360& & 0.36370 \cite{fischer98a}& & 0.36389 \cite{wang94a} & 0.363243\footnotemark[1] \cite{yan98b} &0.35(2) \cite{kernahan75a}  \\
 $2s$$\to$$3p$ & 0.15333& & 0.15346 \cite{fischer98a}& & 0.15376 \cite{qu99b}   & &   0.15(1) \cite{kernahan75a} \\
 $2s$$\to$$4p$ & 0.04969& &                         & & 0.04981 \cite{qu99b}   & &\\
 $2p$$\to$$3s$ & 0.04640& & 0.04636 \cite{fischer98a}& &   & & 0.05(1) \cite{kernahan75a} \\
 $2p$$\to$$3d$ & 0.63801& & 0.63803 \cite{fischer98a}&&&&0.62(6) \cite{kernahan75a}\\
          \multicolumn{8}{c}{B$^{+}$}       \\    
 $2s^2$ $^1$S$^e$$\to$$2s2p$ $^1$P$^o$ &0.99907&1.002 \cite{chen99a} &1.001 \cite{fischer04a} &0.9976(22) \cite{jonsson99a}& 0.9997 \cite{weiss95a}&1.0012\footnotemark[2] \cite{ynnerman95a}&0.98(8) \cite{bashkin85a} \\
 & & &&0.999(5) \cite{godefroid95a}&1.005 \cite{fleming96b}&1.0012\footnotemark[3] \cite{safronova96a}&0.71(5) \cite{kernahan75a}\\
 & & &&&0.9998\footnotemark[4] \cite{savukov04a}&&0.98(6) \cite{irving99a}\\
 & & &&&1.0028\footnotemark[5] \cite{safronova11c}&&                        \\
 $2s^2$ $^1$S$^e$$\to$$2s3p$ $^1$P$^o$ &0.10959&0.108 \cite{chen99a} &0.1087 \cite{fischer04a}&0.1093(3) \cite{fischer97b} &&& \\
 $2s2p$ $^1$P$^o$$\to$$2p^2$ $^1$D$^e$ &0.16195&0.162 \cite{chen99a} &0.1621 \cite{fischer04a}&0.1608(44) \cite{jonsson99a}&0.1625 \cite{weiss95a}&&0.192(9) \cite{bergstrom69a}\\
 &&&&&&&0.114(6) \cite{martinson70a}\\
 $2s2p$ $^1$P$^o$$\to$$2s3d$ $^1$D$^e$ &0.51545 &0.514 \cite{chen99a}&0.5161 \cite{fischer04a}&&0.5199 \cite{weiss95a}&&0.49(2) \cite{kernahan75a} \\
 $2s2p$ $^1$P$^o$$\to$$2p^2$ $^1$S$^e$ &0.22591&0.227 \cite{chen99a}&0.2259 \cite{fischer04a}&0.2257(38) \cite{jonsson99a}&0.2264 \cite{weiss95a} &&0.24(2) \cite{bashkin85a}\\
&&&&&&&0.20(1) \cite{kernahan75a}\\
&&&&&&&0.163(11) \cite{bergstrom69a}\\
 $2s2p$ $^1$P$^o$$\to$$2s3s$ $^1$S$^e$ &0.00008&&0.00019 \cite{fischer04a}&&0.00007 \cite{weiss95a}&&0.039(2) \cite{kernahan75a}\\
 $2s2p$ $^3$P$^o$$\to$$2p^2$ $^3$P$^e$ &0.34298&0.365 \cite{chen05b} &0.34292 \cite{fischer04a}&0.3427(2) \cite{jonsson99a}&0.3427 \cite{weiss95a}&&0.34(3) \cite{bashkin85a}   \\
 &&&&&&&0.32(2) \cite{kernahan75a}\\
 $2s2p$ $^3$P$^o$$\to$$2s3s$ $^3$S$^e$  &0.06377 & &0.06401 \cite{fischer04a}&&&&    \\
 $2s2p$ $^3$P$^o$$\to$$2s3d$ $^3$D$^e$  &0.47627&0.473 \cite{chen99a} &0.47597 \cite{fischer04a} &&&&0.49(2) \cite{kernahan75a}\\
 $2p2p$ $^3$P$^e$$\to$$2p3d$ $^3$D$^o$  &0.62300  & 0.310 \cite{chen05b} &&&&&  \\
\end{tabular}
\footnotetext[1]{Hylleraas$-$type variational method.}
\footnotetext[2]{Multi-Configuration Dirac-Fock (MCDF) method.}
\footnotetext[3]{Relativistic MBPT calculation.}
\footnotetext[4]{Relativistic CI calculation with MBPT theory.}
\footnotetext[5]{Relativistic CI calculation with all order MBPT theory. Calculated with theoretical energy differences.}
\end{ruledtabular}
\end{table*}
\endgroup

The oscillator strengths for the transitions between the low 
manifolds states are listed in Tables \ref{ostrength}. The 
absorption oscillator strength from state $\psi_i$ to state 
$\psi_j$ is calculated according to the identity \cite{yan96a,mitroy03f},  
\begin{equation}
f^{(k)}_{ij} =  \frac {2 |\langle \psi_i;L_i \parallel \  r^k
{\bf C}^{k}({\bf \hat{r}}) \parallel \psi_{j};L_j \rangle|^2 \epsilon_{ji}}
{(2k+1)(2L_i+1)}  \ .
\label{fvaldef}
\end{equation}
In this expression, $\epsilon_{ji} = (E_j - E_i)$ is the energy 
difference between the initial state and final state, while $k$ is 
the multipolarity of the transition, and ${\bf C}^{k}({\bf \hat{r}})$
is a spherical tensor. Experimental energy differences were used 
for the calculation of oscillator strengths.

There have been very many calculations performed of the energy levels 
and oscillator strengths for B$^{2+}$ \cite{fischer98a, wang94a, qu99b, 
yan98b, qu98a, schweizer99a} and B$^+$ \cite{chen99a, chen05b, fischer04a, 
tachiev99a, jonsson99a, litzen98a, fischer97b, godefroid95a, weiss95a, 
hibbert75b, kingston00a, fleming96b, ynnerman95a, ozdemir99a, savukov04a, 
safronova96a}. Not all of the theoretical calculations were tabulated. 
Table \ref{ostrength} gives the reported results of the calculations 
that are deemed to be the most accurate or of particular relevance to 
present calculations.  

For B$^{2+}$ ion $f$-values are given the Multi-Configuration 
Hartree-Fock calculation with Briet-Pauli corrections (MCHF-BP) 
\cite{fischer98a}. The present calculations agree with the 
MCHF-BP values to an accuracy of 0.0001. While the present 
calculations are ostensibly nonrelativistic, they implicitly 
include relativistic corrections since the energies are 
tuned to experimental values. The $2s\to2p$ oscillator strength 
computed with the Hylleraas method \cite{yan98b} is close to the 
non-relativistic limit, but the Hylleraas calculation omits 
any relativistic effects and the Hylleraas energy difference for 
the $2s \to 2p$ transitions is 0.22016 a.u. which is about 0.1$\%$ 
smaller than the experimental energy difference. The full core 
plus correlation calculation \cite{wang94a,qu99b,qu98a} listed in  
the CI column is a variant of the configuration interaction approach.

There is one previous calculation for B$^{+}$ that is very similar 
in concept to the present methodology. That was a CI calculation 
with a semi-empirical core potential \cite{chen99a,chen05b}. The 
major distinction was the adoption of a B-spline basis so this 
calculation is abbreviated as BCICP in Table \ref{ostrength}.
With a few exceptions, the CICP and BCICP oscillator strengths 
agree to about 1$\%$. When the BCICP oscillator strengths are 
different from the present values, one also finds the BCICP 
oscillator strengths also disagreeing with the MCHF-BP B$^{2+}$ 
oscillator strengths \cite{fischer04a}.  

There is also better than 0.3$\%$ agreement of the CICP calculation 
with MCHF oscillator strengths with two exceptions. The MCHF 
oscillator strength \cite{jonsson99a} for the $2s2p$ $^1$P$^o$ $\to$ 
$2p^2$ $^1$D$^e$ transition is about $1\%$ smaller than the CICP 
oscillator strength. The MCHF oscillator strength however is about 
$1\%$ smaller that the BCICP and MCHF-BP oscillator strengths. 
There is also agreement at better than 1$\%$ level with a CI 
calculation \cite{weiss95a} except for the case of the $2s2p$ $^1$P$^o$ 
$\to$ $2s3s$ $^1$S$^e$ transition which has a very small 
oscillator strength.   

There have been two calculations which combine relativistic 
CI calculations with many body perturbation theory (CI+MBPT) 
to represent the core-valence interaction \cite{savukov04a,safronova11c}. 
These only gave the oscillator strength for the $2s^2$ $^1$S$^e$ 
to $2s2p$ $^1$P$^o$ transition. The total range between the CICP 
oscillator strengths and two CI+MBPT oscillator strengths is 
less than 0.4$\%$. The agreement of the CICP oscillator strengths
with another two relativistic calculations that are the MCDF calculation 
of \cite{ynnerman95a} and MBPT calculation of \cite{safronova96a}
is also at 0.2$\%$ level.

Some experimental oscillator strength measurements \cite{irving99a,
bashkin85a,bergstrom69a, martinson70a,kernahan75a} are also listed 
in Table \ref{ostrength} for completeness. The precision of the 
experimental data is not as high as many of the theoretical oscillator
strengths.

\subsection{Scalar and tensor polarizabilities} 

This analysis is done under the premise that spin-orbit
effects are small and the radial parts of the wave functions 
are the same for the states with different $J$.   

All the polarization parameters reported here are calculated using 
their respective oscillator strength sum rules. The multipole 
oscillator strengths $f^{(k)}_{ij}$ are defined in Eq.~(\ref{fvaldef}). 
Then the adiabatic multipole polarizabilities $\alpha_{k}$ from 
the state $i$ are written as \cite{angel68a}
\begin{equation}
\alpha_{k} = \sum_j \frac{f^{(k)}_{ij}}{\epsilon^2_{ji}} \ .
\label{alphav}
\end{equation}
Related sum rules such as the non-adiabatic multipole polarizability 
$\beta_{k}$ and S$_{k}$(-4) are given as \cite{mitroy03f}
\begin{equation}
\beta_{k} = \frac{1}{2}\sum_j \frac{f^{(k)}_{ij}}{\epsilon^3_{ji}} \ ,
\label{betav}
\end{equation}
and \cite{mitroy10a}  
\begin{equation}
S_{k}(-4) = \sum_j \frac{f^{(k)}_{ij}}{\epsilon^4_{ji}} \ .
\label{sum1}
\end{equation}
The $S_{k}(-4)$ sum rule gives the lowest order frequency 
dependent component to the dynamic polarizability through the 
relation  
\begin{equation}
\alpha_{k}(\omega) = \alpha_{k}(0) + \omega^2 S_{k}(-4) + \ldots.  
\label{sum2}
\end{equation}

States with a non-zero angular momentum will also have a tensor 
polarizability \cite{zhang07a,mitroy10a}. For a state with angular 
momentum $L_0$($J_0$), this is defined as the polarizability of the 
magnetic sub-level with $M = L_0$($M = J_0$). The total polarizability 
is written in terms of both a scalar and tensor polarizability.
The scalar polarizability represents the average shift of the
different $M$ levels while the tensor polarizability gives the
differential shift.

This tensor polarizability can be expressed in terms of $f$-value sum 
rules. For an $L_0 = 1$ initial state, one can write the tensor 
polarizability for a dipole field as \cite{mitroy10a}
\begin{equation}
\alpha_{2,L_0L_0} = -\biggl( \ \sum_{n,L_n=0} \frac {f_{0n} } {\epsilon_{n0}^2}
 -\frac{1}{2} \sum_{n,L_n=1} \frac {f_{0n} } {\epsilon_{n0}^2} 
 + \frac{1}{10} \sum_{n,L_n=2} \frac {f_{0n} } {\epsilon_{n0}^2} \biggr) \ .
\label{alpha2L}
\end{equation}
If the initial state is a $L_0 = 2$ state, one can use the expressions in 
\cite{zhang07a} and get the $f$-value sum 
\begin{equation}
\alpha_{2,L_0L_0} = -\biggl( \ \sum_{n,L_n=1} \frac {f_{0n} } {\epsilon_{n0}^2}
        -\sum_{n,L_n=2} \frac {f_{0n} } {\epsilon_{n0}^2}
	+ \frac{2}{7} \sum_{n,L_n=3} \frac {f_{0n} } {\epsilon_{n0}^2} \biggr) \ .
\label{alpha2Lq}
\end{equation}
The core does not make a contribution to the tensor polarizability
since it has an equal impact on all the different $M$-levels.

\begin{table}[t]
\caption[]{ \label{fcore}
The pseudo-oscillator strength distribution for the core B$^{3+}$.
The energy shift parameter $\epsilon_i$ and the adiabatic
($\alpha_{\rm core}$) and non-adiabatic ($\beta_{\rm core}$) core
polarizabilities from Hylleraas calculations \cite{bhatia97a}
are also displayed. The numbers in the square brackets denote powers of 10.}  
\begin{ruledtabular}
\begin{tabular}{lcccc}
\multicolumn{1}{c}{} &  
\multicolumn{1}{c}{$\epsilon_i$}& \multicolumn{1}{c}{$f^{(k)}_i$} & \multicolumn{1}{c}{$\alpha_{\rm core}$}&\multicolumn{1}{c}{$\beta_{\rm core}$} \\
\hline   
dipole & $-$16.67592 &1.0   & 1.9644[$-$2] & 1.1243[$-$3]             \\
& $-$7.89382 &1.0  &   &       \\
quadrupole &$-$21.91592 &0.28537 & 3.4266[$-$3] & 1.5237[$-$4] \\
&  $-$10.14212 & 0.28537   &    &\\
octupole & $-$22.51592    &0.15844 & 1.5216[$-$3]& 5.9751[$-$5]            \\
&$-$11.44722&0.15844&& \\
\end{tabular}
\end{ruledtabular}
\end{table}

The development above is for $LS$ coupled states, but it
is common to give the tensor polarizability for $LSJ$ states.
These can be related to the $LS$ states by geometric factors
arising from the application of Racah algebra. The scalar 
polarizabilities for the different $J$ levels are the same
(if spin-orbit splitting is neglected) and equal to the scalar
polarizability in the $L$ representation. The tensor polarizabilities
between the $L$ and $J$ representations can be related using the 
expressions of \cite{zhang07a}.  
When $L_0 = 1$ and $J_0 = 0$ one finds $\alpha_{2,J_0J_0} = 0$ while 
$J_0 = 1$ case gives $\alpha_{2,J_0J_0} = -\frac{1}{2} \times \alpha_{2,L_0L_0}$.

\begin{table*}[t]
\caption[]{ \label{polara}
The polarizabilities of some low lying states of B$^{2+}$ ion. 
The scalar adiabatic polarizabilities ($\alpha_{k}$) are listed 
along with some non-adiabatic ($\beta_{k}$) and tensor 
($\alpha^{(1)}_{2, L_0 L_0}$ ) polarizabilities. All the 
polarizabilities are calculated using the experimental 
energies. The dipole polarizabilities from accurate CI calculations 
\cite{wang94a,pipin83a} are displayed for comparison.  
The polarizabilities are in atomic units. }  

\begin{ruledtabular}
\begin{tabular}{lcccccc}
\multicolumn{1}{c}{State} &  
\multicolumn{1}{c}{$\alpha_1$ }&\multicolumn{1}{c}{$\beta_1$ }& \multicolumn{1}{c}{$\alpha^{(1)}_{2, L_0 L_0}$ } & \multicolumn{1}{c}{$\alpha_2$ }& \multicolumn{1}{c}{$\beta_2$ }&  \multicolumn{1}{c}{$\alpha_3$} \\
\hline
2s $^2$S$^e$ &  7.8460&17.137& & 7.0963 & 3.8719&    30.181     \\
             &7.847 \cite{wang94a}&&&&&\\
             &7.85 \cite{pipin83a}&&&&&\\
2p $^2$P$^o$ &  $-$0.56938  &6.9896 & 2.1659 & 5.6105  &3.1396  &  48.761     \\
3s $^2$S$^e$ & 182.94   & 1558.1 & & 1539.0 &9096.2 &  14598             \\
3p $^2$P$^o$ & 312.04  &13153&  20.605  & 643.82  &1374.8&  65846       \\
3d $^2$D$^e$ & $-$191.26   &7616.0 & 208.16  & 10.437  &2170.3  & $-$24466 \\
\end{tabular}
\end{ruledtabular}
\end{table*}

\subsubsection{Core polarizabilities} 

The energy distribution of the oscillator strengths originating from 
core excitations was estimated using a semi-empirical technique \cite{mitroy03f}.
This approach utilizes $f$-value sum rules to construct the pseudo-oscillator 
strength distributions,  
\begin{equation}
\alpha_{{\rm k},core} = \sum_{i \in core} \frac {kN_i \langle r^{2k-2}_{i} \rangle} {(\epsilon_i)^2} \ ,
\label{alphacore}
\end{equation}
where $N_i$ is the number of electrons in a core orbital, $\epsilon_i$
is an energy shift parameter. The energy shift parameter was chosen
so that Eq.~(\ref{alphacore}) reproduces accurate estimates of the 
adiabatic and non-adiabatic core polarizabilities determined by close 
to exact calculations for dipole, quadrupole and octupole transitions 
\cite{bhatia97a}.  

The present calculated pseudo-oscillator strength distributions are 
given in Table \ref{fcore}. They can be used in the determination of the 
dynamic polarizabilities and the long range van der Waals coefficients of 
the B$^{2+}$ and B$^{+}$ ions with other atoms.  

\subsubsection{The B$^+$ and B$^{2+}$ polarizabilities} 

Tables \ref{polara} and \ref{polarb} give the scalar adiabatic multipole 
polarizabilities of the lowest five states of the B$^{2+}$ ion and the lowest 
three states of the B$^{+}$ ion. The tensor polarizabilities and non-adiabatic 
polarizabilities as well as the related sum rules $S_{k}(-4)$ of some 
states are also listed. The energies of the lowest lying states 
(i.e. those in Table \ref{energy}) were adjusted to be the same as the 
experimental energies for the polarizability calculations.   

The present CICP dipole polarizability for the B$^{2+}$ ground state is slightly 
smaller than the polarizability of two very accurate CI type calculations 
\cite{pipin83a,wang94a}. The differences do not exceed 0.004 $a_0^3$.
The CI calculations are non-relativistic and are expected to be 
slightly larger than the actual polarizability \cite{tang10a}. 
A comparison for the iso-electronic ion Be$^+$ can be used to 
estimate an uncertainty in the B$^{2+}$ $2s$ states dipole polarizability. 
A previous CICP calculation gave a dipole polarizability of 24.493 
$a_0^3$ which is very close to the recommended value of 24.489(4) 
$a_0^3$ \cite{tang10a}. Assigning an uncertainty of 0.1$\%$ to the B$^{2+}$ 
ground state polarizability would seem to be justified. The uncertainties 
in the polarizabilities for the excited states are expected to be 
of the same order as that of the ground state except for the case of 
the $2p$ state where considerable cancellations occur in the 
oscillator strength sum rule.

\begin{table*}[t]
\caption[]{ \label{polarb}
The polarizabilities of the lowest three states of the B$^+$ ion. 
The scalar adiabatic polarizabilities ($\alpha_{k}$) are listed along 
with some non-adiabatic ($\beta_{k}$) and tensor ($\alpha^{(1)}_{2, L_0 L_0}$ ) 
polarizabilities. Values for sum rules, $S_{1}$($-$4),  
are also presented. All these values are calculated using the experimental 
energies. The dipole polarizabilities from the relativistic calculation of 
\cite{safronova11c} are displayed for comparison. All polarizabilities 
are in atomic units. }  

\begin{ruledtabular}
\begin{tabular}{lccccccc}
\multicolumn{1}{c}{State} &  
\multicolumn{1}{c}{$\alpha_1$ }&\multicolumn{1}{c}{$S_{1}$($-$4) }&\multicolumn{1}{c}{$\beta_1$ }& \multicolumn{1}{c}{$\alpha^{(1)}_{2, L_0 L_0}$ } & \multicolumn{1}{c}{$\alpha_2$ }& \multicolumn{1}{c}{$\beta_2$ }&  \multicolumn{1}{c}{$\alpha_3$} \\
\hline
$2s^2$ $^1$S$^e$& 9.6442 &80.891 &13.757& &27.138  &20.631&  147.01  \\
          & 9.624 CI+all \cite{safronova11c}&&&&&&\\
$2s2p$ $^3$P$^o$&7.7798&66.320 &10.737&1.4613 & 25.011 &17.940&  220.98  \\
          &7.772 CI+all \cite{safronova11c}&&&&&&\\
$2s2p$ $^1$P$^o$&16.554&603.57 &54.587&$-$2.1960 &44.757  &44.266& 541.44 \\                
\end{tabular}
\end{ruledtabular}
\end{table*}

Table \ref{polarb} gives the polarizabilities of the B$^+$ states. 
The only other calculation of the polarizabilities for these states 
is a recent CI+MBPT calculation \cite{safronova11c}. The CI+MBPT 
calculation gave a polarizability for the $2s^2$ $^1$S$^e$ ground 
state that is 0.2$\%$ smaller than the present CICP calculation.
The difference for the $2s2p$ $^3$P$^o$ state is 0.1$\%$.  

A rough estimate of the uncertainties in the B$^{+}$ polarizabilities 
is possible by reference to similar calculations for the Si$^{2+}$ 
ground state \cite{mitroy08k}. A CICP calculation gave 11.688 $a_0^3$, 
a revised analysis of a resonant excitation Stark ionization 
spectroscopy (RESIS) experiment gave 11.669 $a_0^3$ 
\cite{komara05a,mitroy08k}, and a CI+MBPT calculation gave 11.670(13) 
$a_{0}^3$ \cite{safronova12e}. The comparison between the RESIS
and the CICP polarizabilities suggested that a conservative 
estimate of the ground state $3s^2$ $^1$S$^e$ polarizability was 
0.25$\%$ while that for the excited states was 0.5$\%$. 

Comparisons for Al$^+$ between CICP polarizabilities \cite{mitroy09b} and 
CI+MBPT calculations \cite{safronova11c} reveal differences between 
the two calculations that do not exceed 0.4$\%$. The CI+MBPT 
calculation uses theoretical differences in the calculation of 
the Al$^+$ polarizabilities and overestimates the $3s^2$ $^1$S$^e$ 
- $3s3p$ $^3$P$^o_0$ energy difference by 0.14$\%$. The replacement 
of the theoretical energy differences by the experimental energy 
differences would reduce the difference between the CICP and CI+MBPT 
calculation to less than 0.3$\%$. The analysis for the Al$^+$ system 
suggests that an uncertainty of 0.3$\%$ should be assigned to the 
polarizability of the $2s^2$ $^1$S$^e$ state. So the final recommended 
$2s^2$ $^1$S$^e$ dipole polarizability is 9.64(3) $a_0^3$. Assuming the 
$2s2p$ $^3$P$^o_0$ state has the same uncertainty, the final CICP 
dipole polarizability is 7.78(3) $a_0^3$. The scalar dipole 
polarizability for $2s2p$ $^1$P$^o$ state was 16.55(5) $a_0^3$
assuming the same relative uncertainty.

Table \ref{alphatable} gives a breakdown of the different contributions 
to the $2s^2$ $^1$S$^e$, $2s2p$ $^3$P$^o_0$, and $2s2p$ $^1$P$^o$ dipole 
polarizabilities. About 90$\%$ of the $2s^2$ $^1$S$^e$ polarizability 
comes from the resonant transition and much of the difference with 
the CI+MBPT calculation comes from this transition with the CI+MBPT 
calculation giving 8.918 $a_0^3$ \cite{safronova11c}. The CI+MBPT 
calculation overestimated the $2s^2$ $^1$S$^e$ - $2s2p$ $^1$P$^o$ 
transition energy difference by 0.27$\%$. So it is possible that part 
of the discrepancy with the CI+MBPT calculation could be removed by 
using the experimental energy difference when calculating the 
polarizability of the resonant transition.  It should be noted that 
experimental energy differences were used in a recent CI+MBPT calculation 
of the polarizability of the $3s^2$ state of Si$^{2+}$ \cite{safronova12e}.  

\begin{table}[t]
\caption[]{ \label{alphatable}
Breakdown of the contributions to the dipole polarizabilities of the B$^{+}$
clock transition states. The $\delta \alpha_1$ column gives the contribution
from the indicated transition class. The $\sum \alpha_1$ column gives
the accumulated sum. The final polarizabilities are given in bold-face.
}
\begin{ruledtabular}
\begin{tabular}{lcc}
\multicolumn{1}{c}{Transition(s)}& \multicolumn{1}{c}{$\delta \alpha_1$} & \multicolumn{1}{c}{$\sum$ $\alpha_1$} \\
\hline   
\multicolumn{3}{c}{$2s^2$ $^1$S$^e$ state}    \\
$2s^2$ $^1$S$^e$ $ \to$ $2s2p$ $^1$P$^o$ & 8.9333    & 8.9333     \\
$2s^2$ $^1$S$^e$ $\to$ $2s3p$ $^1$P$^o$  & 0.2542   & 9.1875     \\
$2s^2$ $^1$S$^e$ $\to$ $nP$ $^1$P$^o$    & 0.4370     & 9.6245     \\
Core                                     &  0.01964    &  {\bf 9.6441}  \\
\multicolumn{3}{c}{$2s2p$ $^3$P$^o_0$ state}    \\
$2s2p$ $^3$P$^o$ $\to$ $2s3s$ $^3$S$^e$    & 0.3596  &  0.3596\\
$2s2p$ $^3$P$^o$ $\to$ $nS$ $^3$S$^e$      & 0.1093  &  0.4689 \\
$2s2p$ $^3$P$^o$ $\to$ $2p^2$ $^3$P$^e$    & 4.3573  &  4.8262  \\
$2s2p$ $^3$P$^o$ $\to$ $nP$ $^3$P$^e$      & 0.0750  &  4.9012     \\
$2s2p$ $^3$P$^o$ $\to$ $2s3d$ $^3$D$^e$    & 1.7873  &  6.6885    \\
$2s2p$ $^3$P$^o$ $\to$ $nD$ $^3$D$^e$      & 1.0717  &  7.7602   \\
Core                            &  0.01964   & {\bf 7.7798 }    \\
\multicolumn{3}{c}{$2s2p$ $^1$P$^o$ state}    \\
$2s2p$ $^1$P$^o$ $\to$ $2s^2$ $^1$S$^e$    & $-$2.9778  &$-$2.9778\\
$2s2p$ $^1$P$^o$ $\to$ $2p^2$ $^1$S$^e$    & 3.6955  &0.7177\\
$2s2p$ $^1$P$^o$ $\to$ $nS$ $^1$S$^e$      & 0.0707 &  0.7884\\
$2s2p$ $^1$P$^o$ $\to$ $2p^2$ $^1$D$^e$    & 9.2975  &  10.0859  \\
$2s2p$ $^1$P$^o$ $\to$ $2s3d$ $^1$D$^e$    & 3.7574  &  13.8433  \\
$2s2p$ $^1$P$^o$ $\to$ $nD$ $^1$D$^e$      & 2.4125  & 16.2558      \\
$2s2p$ $^1$P$^o$ $\to$ $2p3p$ $^1$P$^e$    & 0.1613  & 16.4171    \\
$2s2p$ $^1$P$^o$ $\to$ $nP$ $^1$P$^e$      & 0.1169  & 16.5340    \\
Core                                       &  0.01964 & {\bf 16.5536}    \\
\end{tabular}
\end{ruledtabular}
\end{table}

The present CICP calculation of the ground state polarizability does not 
take into consideration the contribution from the $2s^2$ $^1$S$^e$ $\to$ 
$2s2p$ $^3$P$^o_1$ transition. The oscillator strength for this transition 
is only $3.361 \times 10^{-8}$ \cite{fischer04a} so this transition can 
be safely omitted from the determination of the  polarizability.  This 
also justifies the omission of the spin-orbit interaction from the 
effective Hamiltonian for the valence electrons.  

\subsection{The BBR shift}

The blackbody radiation shift of an atomic clock transition can be 
approximately calculated using the Eqs.~(\ref{BBR3}) and (\ref{BBR1}). 
In this expression the temperature in K is multiplied by 3.1668153 
$\times$ 10$^{-6}$. Using the present polarizabilities and converting 
to frequency shifts at 300 K gives 
$\Delta \nu_{2s^2 \ ^1\text{S}^e} =$ $-$0.08305 Hz and 
$\Delta \nu_{2s2p \ ^3\text{P}^o_0} =$ $-$0.06699 Hz. 
In the present CICP calculation the dipole polarizability difference 
for the $2s^2$ $^1$S$^e$ $\to$ $2s2p$ $^3$P$^o_0$ clock transition is 
$\Delta \alpha_1=$ $-$1.8643 $a_0^3$. The relativistic CI+MBPT 
calculation \cite{safronova11c} gave $\Delta \alpha_1=$ $-$1.851 $a_0^3$. 

Using a value of $\Delta \alpha_1=$ $-$1.8643 $a_0^3$ leads to 
a net frequency shift at 300 K of $\Delta \nu= 0.01605$ Hz. This 
is consistent with the CI+MBPT result $\Delta \nu= 0.0159(16)$ Hz 
\cite{safronova11c}. A small correction to the polarizabilities 
needs to be made to allow for the slight variation of the 
polarizabilities due to the finite temperature of the BBR radiation 
field,    
\begin{equation}
\alpha_1(T)=\alpha_1(1+\eta),
\end{equation}
where $\alpha_1(T)$ is the polarizability after correction. $\eta$ 
is the dynamic correction factor. The leading order term of $\eta$ 
is given by \cite{porsev06a}
\begin{equation}
\eta \approx  -\frac{40 \pi^2 T^2}{21 \alpha_1(0)} S_1(-4) \ .
\label{eta}
\end{equation}
The value of $\eta$ was found to be quite small. In the present CICP 
calculation, it was $-$$1.42 \times 10^{-4}$ for the $2s^2$ $^1$S$^e$
state and $-$$1.45 \times 10^{-4}$ for the $2s2p$ $^3$P$^o_0$ state.
Taking this correction into account, the 300 K dipole polarizabilities
of $2s^2$ $^1$S$^e$ state and $2s2p$ $^3$P$^o$ state are 9.6428 $a_0^3$
and 7.7787 $a_0^3$ respectively. The polarizability difference is 
$\Delta \alpha_1= -1.8642$ $a_0^3$. This is only 0.0001 $a_0^3$ smaller
than the $T = 0$ K value of $-$1.8643 $a_0^3$. It is evident that the
effect of the dynamic correction in the B$^+$ $2s^2$ $^1$S$^e$ $\to$
$2s2p$ $^3$P$^o_0$ clock transition is miniscule.

When the uncertainties in the polarizabilities are taken into 
consideration the final recommended CICP polarizability difference at 
300 K is $-$1.86(6) $a_0^3$. The derived frequency shift is 0.0160(5) Hz. 
The CI+MBPT calculation gave a polarizability difference of $-$1.85(18) 
$a_0^3$ and a frequency shift of 0.0159(16) Hz. The difference between the 
CICP and CI+MBPT calculations of the frequency shift is less than 
1.0$\%$. 

The uncertainty associated with the CI+MBPT calculation is more than 
three times larger than that quoted for the present CICP calculation.  
Although uncertainties are not assigned to the CI+MBPT polarizabilities, 
their final BBR shift uncertainty indicates uncertainties in their 
polarizabilities of 1.0$\%$.  The CI+MBPT uncertainty estimates seem 
very conservative given the 0.02 $a_0^3$ level of agreement between the 
CICP and CI+MBPT polarizabilities.  A more recent CI+MBPT calculation of 
the polarizability of the Si$^{2+}$ ground state quoted an 
uncertainty of 0.12$\%$ \cite{safronova12e}.  

\section{Conclusions}

The polarizabilities of some low lying states of the B$^{2+}$ 
and B$^+$ ions are computed with large basis CI calculations with 
an underlying semi-empirical Hamiltonian. The motivation for 
these calculations was an independent calculation of the BBR 
shift of the B$^+$ $2s^2$ $^1$S$^e$ $\to$ $2s2p$ $^3$P$^{o}_0$ 
clock transition \cite{safronova11c}. 

The final estimate of the frequency shift, namely 0.0160(5) Hz 
is within 1$\%$ of the earlier CI+MBPT calculations \cite{safronova11c}.
The almost perfect agreement between these two completely independent
calculations gives increased confidence in the respective reliabilities
of both calculations. One reason for the good agreement between both
calculations is that both calculations give very accurate solutions
of the Schrodinger equation with respect to their underlying Hamiltonian.
Both the CICP and CI+MBPT approximate the aspects of the physics,
and in particular the core-valence interaction. The CICP calculation 
uses a HF plus semi-empirical polarization potential to simulate
core-valence correlation effects. The CI+MBPT calculation uses MBPT
to incorporate the dynamical effects going beyond the HF interaction.
Making these approximations simplifies the calculation sufficiently
to allow a close to numerically exact solutions of the Schrodinger
equation for the two valence electrons. 

In addition to the dipole polarizabilities, the present model computes 
the quadrupole, octupole and non-adiabatic dipole polarizabilities.    
One way to measure the B$^+$ polarizability would be the RESIS technique 
\cite{lundeen05a,snow08a}.  
The analysis of the raw experimental RESIS data can be improved if estimates 
of the quadrupole and non-adiabatic dipole polarizability are 
available.  

\begin{acknowledgments}

The authors would like to thank Dr Marianna Safronova for helpful 
communications and sharing unpublished data. The work of J.M was 
supported by the Australian Research Council Discovery Project DP-1092620.
Dr Yongjun Cheng was supported by a grant from the Chinese Scholarship 
Council. 

\end{acknowledgments}



\end{document}